\documentclass[twocolumn,english,preprintnumbers,amsmath,amssymb,superscriptaddress]{revtex4}
\usepackage[latin9]{inputenc}
\usepackage{amsmath}
\usepackage{graphicx}

\makeatletter
\@ifundefined{textcolor}{}
{%
 \definecolor{BLACK}{gray}{0}
 \definecolor{WHITE}{gray}{1}
 \definecolor{RED}{rgb}{1,0,0}
 \definecolor{GREEN}{rgb}{0,1,0}
 \definecolor{BLUE}{rgb}{0,0,1}
 \definecolor{CYAN}{cmyk}{1,0,0,0}
 \definecolor{MAGENTA}{cmyk}{0,1,0,0}
 \definecolor{YELLOW}{cmyk}{0,0,1,0}
}

\usepackage{dcolumn}
\usepackage{bm}
\usepackage{color}
\usepackage[final]{pdfpages}

\makeatother

\usepackage{babel}

\begin{document}

\preprint{preprint(\today)}

\title{Charge orders with distinct magnetic response in a prototypical kagome superconductor LaRu$_{3}$Si$_{2}$}

\author{C.~Mielke III}
\thanks{These authors contributed equally to the paper.}
\affiliation{Laboratory for Muon Spin Spectroscopy, Paul Scherrer Institute, CH-5232 Villigen PSI, Switzerland}
\affiliation{Physik-Institut, Universit\"{a}t Z\"{u}rich, Winterthurerstrasse 190, CH-8057 Z\"{u}rich, Switzerland}

\author{V.~Sazgari}
\thanks{These authors contributed equally to the paper.}
\affiliation{Laboratory for Muon Spin Spectroscopy, Paul Scherrer Institute, CH-5232 Villigen PSI, Switzerland}

\author{I.~Plokhikh}
\affiliation{Laboratory for Multiscale Materials Experiments, Paul Scherrer Institut, 5232, Villigen PSI, Switzerland}

\author{S.~Shin}
\affiliation{Laboratory for Multiscale Materials Experiments, Paul Scherrer Institut, CH-5232 Villigen PSI, Switzerland}

\author{H.~Nakamura}
\affiliation{Institute for Solid State Physics (ISSP), University of Tokyo, Kashiwa, Chiba 277-8581, Japan}

\author{J.N.~Graham}
\affiliation{Laboratory for Muon Spin Spectroscopy, Paul Scherrer Institute, CH-5232 Villigen PSI, Switzerland}

\author{J.~K\"{u}spert}
\affiliation{Physik-Institut, Universit\"{a}t Z\"{u}rich, Winterthurerstrasse 190, CH-8057 Z\"{u}rich, Switzerland}

\author{I.~Bia\l{}o}
\affiliation{Physik-Institut, Universit\"{a}t Z\"{u}rich, Winterthurerstrasse 190, CH-8057 Z\"{u}rich, Switzerland}

\author{G.~Garbarino}
\affiliation{European Synchrotron Radiation Facility, 71 Avenue des Martyrs, 38000 Grenoble, France}

\author{D.~Das}
\affiliation{Laboratory for Muon Spin Spectroscopy, Paul Scherrer Institute, CH-5232 Villigen PSI, Switzerland}

\author{M.~Medarde}
\affiliation{Laboratory for Multiscale Materials Experiments, Paul Scherrer Institut, CH-5232 Villigen PSI, Switzerland}

\author{M.~Bartkowiak}
\affiliation{Laboratory for Neutron and Muon Instrumentation, Paul Scherrer Institut, CH-5232 Villigen, Switzerland}

\author{S.S.~Islam}
\affiliation{Laboratory for Muon Spin Spectroscopy, Paul Scherrer Institute, CH-5232 Villigen PSI, Switzerland}

\author{R.~Khasanov}
\affiliation{Laboratory for Muon Spin Spectroscopy, Paul Scherrer Institute, CH-5232 Villigen PSI, Switzerland}

\author{H.~Luetkens}
\affiliation{Laboratory for Muon Spin Spectroscopy, Paul Scherrer Institute, CH-5232 Villigen PSI, Switzerland}

\author{M.Z. Hasan}
\affiliation{Laboratory for Topological Quantum Matter and Advanced Spectroscopy (B7), Department of Physics,
Princeton University, Princeton, New Jersey 08544, USA}

\author{E.~Pomjakushina}
\affiliation{Laboratory for Multiscale Materials Experiments, Paul Scherrer Institut, 5232, Villigen PSI, Switzerland}

\author{J.-X.~Yin}
\affiliation{Department of Physics, Southern University of Science and Technology, Shenzhen, Guangdong, 518055, China}

\author{M.~H.~Fischer}
\affiliation{Physik-Institut, Universit\"{a}t Z\"{u}rich, Winterthurerstrasse 190, CH-8057 Z\"{u}rich, Switzerland}

\author{J.~Chang}
\affiliation{Physik-Institut, Universit\"{a}t Z\"{u}rich, Winterthurerstrasse 190, CH-8057 Z\"{u}rich, Switzerland}

\author{T. Neupert}
\affiliation{Physik-Institut, Universit\"{a}t Z\"{u}rich, Winterthurerstrasse 190, CH-8057 Z\"{u}rich, Switzerland}

\author{S.~Nakatsuji}
\affiliation{Institute for Solid State Physics, University of Tokyo, Kashiwa, 277-8581, Japan}

\author{B.~Wehinger}
\affiliation{European Synchrotron Radiation Facility, 71 Avenue des Martyrs, 38000 Grenoble, France}

\author{D.J.~Gawryluk}
\email{dariusz.gawryluk@psi.ch} 
\affiliation{Laboratory for Multiscale Materials Experiments, Paul Scherrer Institut, 5232, Villigen PSI, Switzerland}

\author{Z.~Guguchia}
\email{zurab.guguchia@psi.ch} 
\affiliation{Laboratory for Muon Spin Spectroscopy, Paul Scherrer Institute, CH-5232 Villigen PSI, Switzerland}

\begin{abstract}

\end{abstract}

\maketitle

\textbf{The kagome lattice\cite{Syozi,JiaxinNature,GuguchiaCSS,Kiesel} has emerged as a promising platform for hosting unconventional chiral charge order at high temperatures. Notably, in LaRu$_{3}$Si$_{2}$, a room-temperature charge-ordered state with a propagation vector of ($\frac{1}{4}$,~0,~0) has been recently identified \cite{GuguchiaPlokhikh}. However, understanding the interplay between this charge order and superconductivity, particularly with respect to time-reversal-symmetry breaking, remains elusive. In this study, we employ single crystal X-ray diffraction, magnetotransport, and muon-spin rotation experiments to investigate the charge order and its electronic and magnetic responses in LaRu$_{3}$Si$_{2}$ across a wide temperature range down to the superconducting state. Our findings reveal the emergence of a charge order with a propagation vector of ($\frac{1}{6}$,~0,~0) below $T_{\rm CO,2}$ ${\simeq}$ 80 K, coexisting with the previously identified room-temperature primary charge order ($\frac{1}{4}$,~0,~0). The primary charge-ordered state exhibits zero magnetoresistance. In contrast, the appearance of the secondary charge order at $T_{\rm CO,2}$ is accompanied by a notable magnetoresistance response and a pronounced temperature-dependent Hall effect, which experiences a sign reversal, switching from positive to negative below $T^{*}$ ${\simeq}$ 35 K. Intriguingly, we observe an enhancement in the internal field width sensed by the muon ensemble below $T^{*}$ ${\simeq}$ 35 K. Moreover, the muon spin relaxation rate exhibits a substantial increase upon the application of an external magnetic field below $T_{\rm CO,2}$ ${\simeq}$ 80 K. Our results highlight the coexistence of two distinct types of charge order in LaRu$_{3}$Si$_{2}$ within the correlated kagome lattice, namely a non-magnetic charge order ($\frac{1}{4}$,~0,~0) below $T_{\rm co,1}$ ${\simeq}$ 400 K and a time-reversal-symmetry-breaking charge order below $T_{\rm CO,2}$, coexisting with superconductivity. This study sheds light on the intricate electronic and magnetic phenomena occurring in kagome superconductors, providing valuable insights into their unique properties and potential applications.}


\begin{figure*}[t!]
\centering
\includegraphics[width=1.45\linewidth]{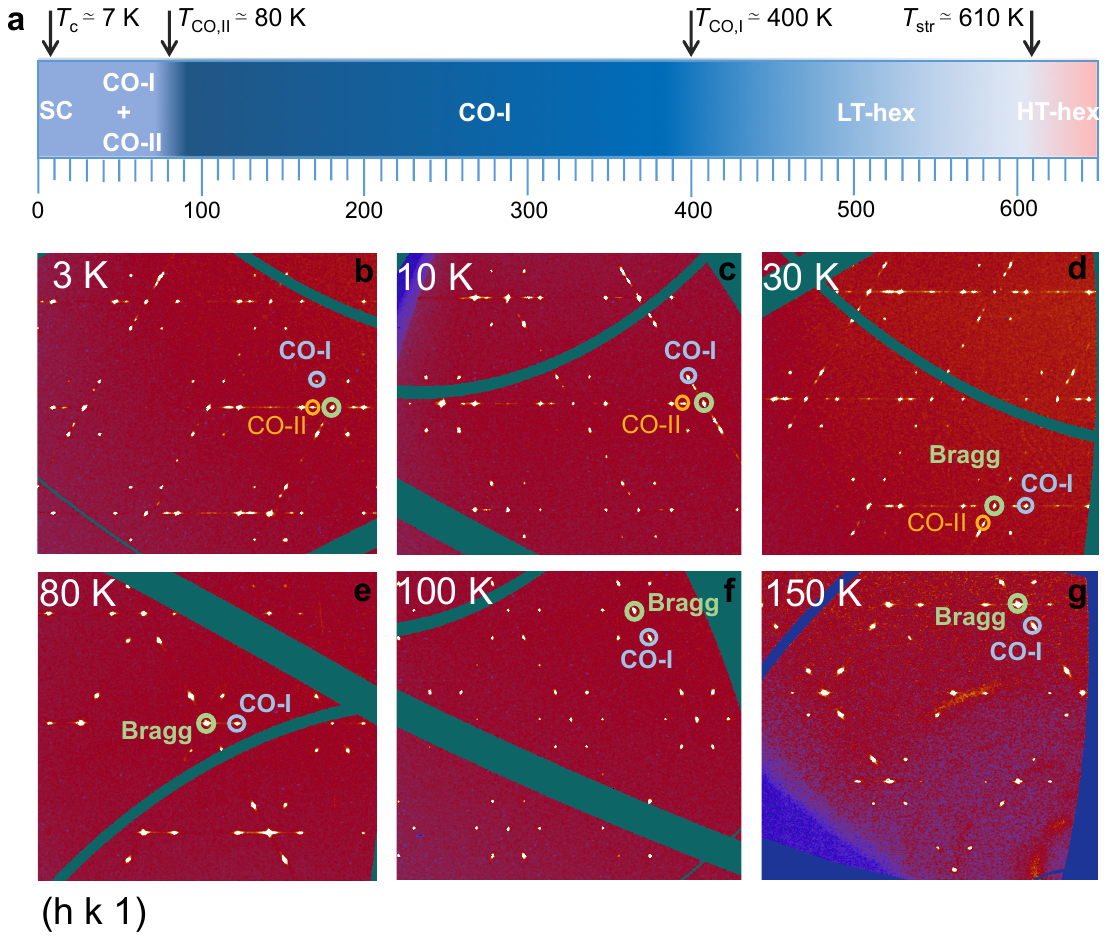}
\vspace{-3.3cm}
\caption{\textbf{Charge order in LaRu$_{3}$Si$_{2}$.} 
$\bf{a,}$ Schematic phase diagram as a function of temperature of LaRu$_{3}$Si$_{2}$. The arrows mark the structural phase transition temperature $T_{\rm str}$ from the high-temperature hexagonal $P6/mmm$ phase (SG No.~191) to the low-temperature orthorhombic $Cccm$ phase (SG No.~66), primary charge order transition temperature $T_{\rm CO-I}$ and secondary charge order transition temperature $T_{\rm CO-II}$. $\bf{b-g,}$ Reconstructed reciprocal space along the (h~k~1) direction, performed at various temperatures above and below the superconducting transition temperature.}
\label{fig1}
\end{figure*}

Kagome superconductors \cite{Barz,Vandenberg,GuguchiaPRM,BOrtiz2,BOrtiz3,QYin}, distinguished by their unique lattice structure resembling the traditional Japanese kagome pattern, exhibit diverse electronic behaviors  and have become a focal point in the study of condensed matter physics. One intriguing aspect is their propensity to host chiral charge order \cite{YJiang,GuguchiaMielke,GuguchiaNPJ,TNeupert,MHChristensen2022,GWagner,Grandi}, a phenomenon where the electron density is spatially modulated to give the system a handedness, introducing unconventional electronic states. Understanding this interrelationship between kagome superconductivity and chiral charge order is pivotal for harnessing their unique properties in the development of novel quantum materials and technologies.

Four distinct classes of kagome lattice systems have recently been identified as exhibiting charge order: the $A$V$_{3}$Sb$_{5}$ family (where $A$ = K, Rb, Cs) \cite{BOrtiz2,BOrtiz3,QYin}, ScV$_{6}$Sn$_{6}$ \cite{Arachchige,GuguchiaSc166}, FeGe \cite{JiaxinPRL,XTeng} and LaRu$_{3}$Si$_{2}$ \cite{GuguchiaPlokhikh}. In $A$V$_{3}$Sb$_{5}$, a correlated interplay of charge order emerges below $T_{\rm co}$ (around 80-110 K), alongside a superconducting phase developing below $T_{\rm c}$ (approximately 1-3 K). ScV$_{6}$Sn$_{6}$, sharing a similar vanadium structural motif with $A$V$_{3}$Sb$_{5}$, displays charge order below $T_{\rm co}$ (around 90 K) but does not exhibit superconductivity down to the lowest measured temperatures. In contrast, FeGe, characterized as a correlated magnetic kagome system, demonstrates A-type antiferromagnetic order below 400 K and charge order below 100 K. One of the most remarkable features of the charge ordered state in all three materials is the emergence of possible time-reversal symmetry (TRS)-breaking, featuring both magnetic and electronic anomalies \cite{YJiang,GuguchiaMielke,GuguchiaNPJ,TNeupert,MHChristensen2022,GWagner,GuguchiaRVS,NShumiya,Wang2021,KhasanovCVS,LiYu,YXu,GuoMoll,SYang,FYu,MDenner,MHChristensen,ERitz,Tazai2022,Balents,Nandkishore,Qimiao}. Theoretical considerations suggest that these features may be explained by a complex order parameter, realizing a higher angular-momentum state, akin to unconventional superconducting orders. LaRu$_{3}$Si$_{2}$ stands out within the realm of bulk kagome-lattice superconductors due to its exceptional characteristics. Notably, it boasts the highest superconducting critical temperature, reaching approximately 7 K \cite{GuguchiaPRM,Kishimoto,LiWen,LiZeng,YWang}. Furthermore, using single crystal X-ray diffraction we recently showed that LaRu$_{3}$Si$_{2}$ exhibits a charge-ordered state at temperatures above room temperature \cite{GuguchiaPlokhikh}. The coexistence of both high-temperature charge order and notable superconductivity makes LaRu$_{3}$Si$_{2}$ a particularly intriguing system for the exploration of quantum phenomena and potential applications in the field of condensed matter physics.


\begin{figure*}[t!]
\centering
\includegraphics[width=1.0\linewidth]{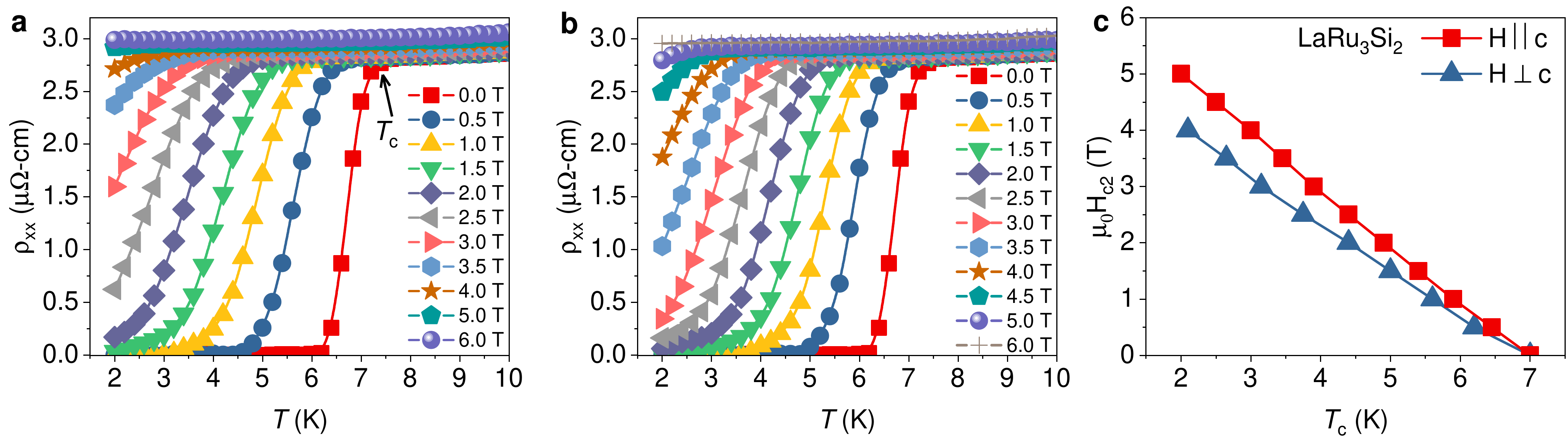}
\vspace{-0.7cm}
\caption{\textbf{Weakly anisotropic superconducting properties of LaRu$_{3}$Si$_{2}$.} 
The temperature dependence of the electrical resistivity of a single crystal of LaRu$_{3}$Si$_{2}$, measured under various magnetic fields applied perpendicular to the $c$-axis ($\mathbf{a}$) and parallel to the $c$-axis ($\mathbf{b}$). $\mathbf{c}$, the magnetic field-temperature phase diagram is derived from the data shown in panels $\mathbf{a}$ and $\mathbf{b}$.}
\label{fig1}
\end{figure*}

X-ray diffraction measurements on LaRu$_{3}$Si$_{2}$ have previously \cite{GuguchiaPlokhikh} been constrained to temperatures above 80 K, leaving uncertainties regarding the persistence of charge order into the superconducting state. Additionally, a comprehensive understanding of the electronic and magnetic response in the normal state remained elusive. Here, we address these questions to gain insights into the charge order across a broad temperature range, encompassing the superconducting state, as well as to elucidate the electronic and magnetic behaviors in the normal state. We employ a set of techniques, including single crystal X-ray diffraction, magnetotransport measurements, and a combination of zero-field and high-field muon spin relaxation/rotation (${\mu}$SR) experiments on a single crystal of LaRu$_{3}$Si$_{2}$. This multifaceted experimental approach aims to provide a comprehensive understanding of the charge-order phenomena and associated electronic and magnetic responses in both the normal and superconducting states of LaRu$_{3}$Si$_{2}$. Studying the magnetic response of charge orders is crucial for understanding the underlying quantum phenomena and the role they play in the emergent properties of these materials.

In Figures 1$\bf{b-g}$, we present reconstructed reciprocal-space patterns along the (hk1) direction for various temperatures both above and below the superconducting critical temperature $T_{\rm c}$, which is approximately 7 K, for a single crystal of LaRu$_{3}$Si$_{2}$. At a higher temperature, $T$ = 150 K, the diffraction pattern reveals fundamental Bragg peaks ${\tau}$ and superlattice peaks at $Q$ = ${\tau}$ + $q_{i}$ with $q_{1}$ = ($\frac{1}{4}$,0,0), $q_{2}$ = (0,$\frac{1}{4}$,0) and $q_{3}$ = ($\frac{1}{4}$,-$\frac{1}{4}$,0). As the temperature decreases below $T_{\rm CO,2}$, approximately 80 K, an additional set of reflections emerges at positions corresponding to $q_{1}'$=($\frac{1}{6},0,0$), $q_{2}'$=(0,~$\frac{1}{6},0$), and $q_{3}'$=($\frac{1}{6},\frac{-1}{6},~0$). Significantly, both $\frac{1}{4}$ and $\frac{1}{6}$ charge orders coexist below 80 K, persisting into the superconducting state. Figure 1$\bf{a}$ illustrates a schematic diagram showing the evolution of structural and electronic properties with respect to temperature. The diagram captures the progression of the material through a structural phase transition from the high-temperature hexagonal $P6/mmm$ phase (SG No.~191) to the low-temperature orthorhombic $Cccm$ phase (SG No.~66) at $T_{\rm str}$ ${\simeq}$ 600 K. Subsequently, it highlights the onset of primary ($\frac{1}{4}$) charge order at $T_{\rm co,1}$ ${\simeq}$ 400 K and the emergence of secondary ($\frac{1}{6}$) charge order below $T_{\rm co, 2}$ ${\simeq}$ 80 K. Notably, both charge orders coexist with superconductivity, manifesting below 7 K.



The superconducting transitions in LaRu$_{3}$Si$_{2}$ were examined in electrical resistivity experiments with the magnetic field applied both perpendicular (Fig. 2$\bf{a}$) and parallel (Fig. 2$\bf{b}$) to the $c$ axis. The results indicate a weak anisotropy of the second critical field $H_{c2}$. The estimated anisotropy of $H_{c2}$ is approximately 1.25, a value significantly lower than the observed anisotropy of 8 in CsV$_{3}$Sb$_{5}$. In our investigation of the electronic properties of LaRu$_{3}$Si$_{2}$ in its normal state, comprehensive macroscopic magnetization and resistivity measurements were conducted under the influence of an applied magnetic field. Figure 3$\bf{a}$ illustrates the temperature dependence of magnetic susceptibility at a magnetic field strength of 1 T, both parallel and perpendicular to the $c$ axis. When the magnetic field is applied parallel to the $c$ axis ($H$ ${\parallel}$ $c$), three distinct anomalies in the magnetic susceptibility are observed. Initially, a subtle drop in susceptibility occurs around 350 K, closely associated with the onset of the primary charge order at $T_{\rm co,1}$. Subsequently, a sharp increase is evident below the secondary charge-order temperature $T_{\rm CO,2}$, approximately 80 K, followed by an additional notable increase below $T^{*}\approx 35$ K. Interestingly, these magnetic susceptibility anomalies are prominently observed when the magnetic field is aligned along the $c$ axis, but they are significantly less discernible when the field is applied in the kagome plane. In the case of $H$ ${\perp}$ $c$, only the transition at $T^{*}$ is evident. This indicates that the magnetic response across the charge-order temperatures is anisotropic. The corresponding temperatures are highlighted in Figure 3$\bf{b}$, where the temperature-dependent resistivity under zero field, as well as its derivative, are presented. The derivative of the resistivity exhibits a peak at $T^{*}$, while a subtle change in the slope of the resistivity curve is observed across the charge ordering temperatures $T_{\rm co,1}$ and $T_{\rm CO,2}$. 


\begin{figure*}[t!]
\centering
\includegraphics[width=1.0\linewidth]{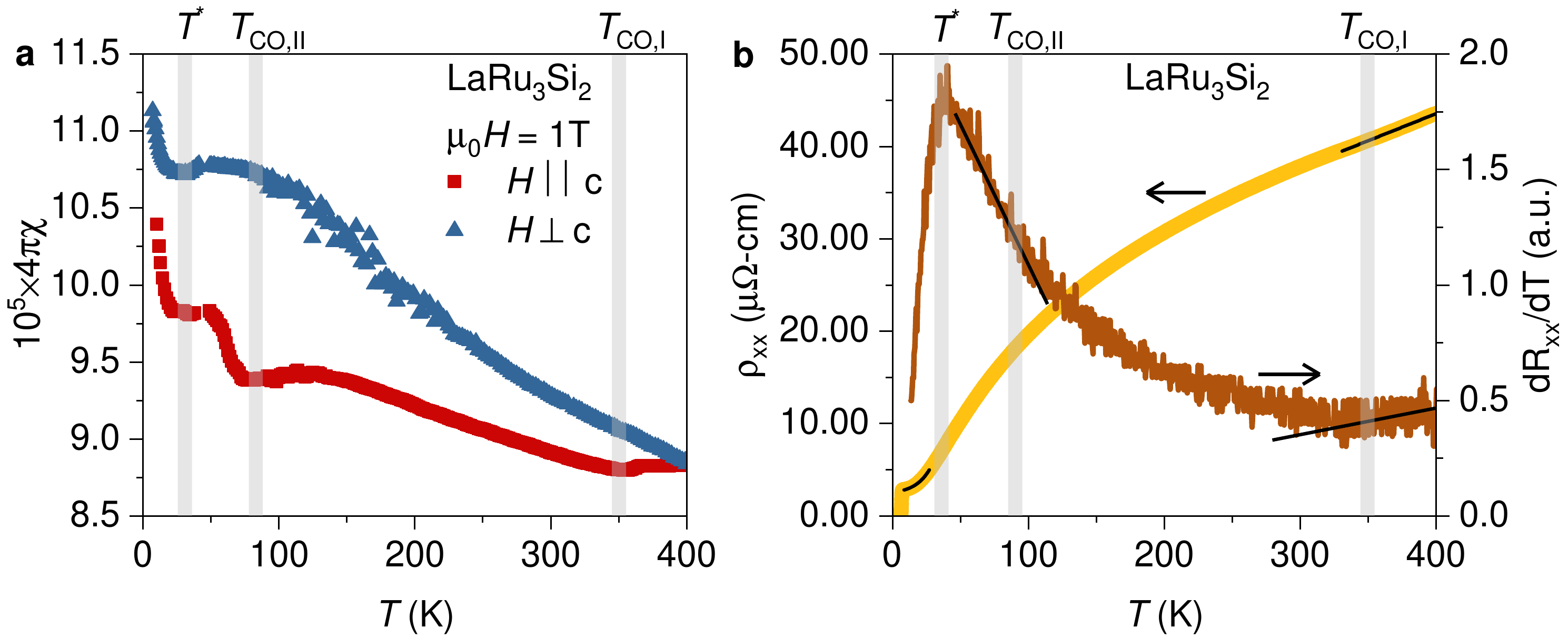}
\vspace{-0.7cm}
\caption{\textbf{Magnetic and transport anomalies in LaRu$_{3}$Si$_{2}$.} 
$\bf{a,}$ The temperature dependence of magnetic susceptibility, measured under the magnetic field of 1 T, applied parallel and perpendicular to $c$-axis. $\bf{b,}$ Temperature dependence of the normal state longitudinal resistance ${\rho}_{xx}$ and its first derivative. Vertical grey lines mark three characteristic temperatures $T^{*}$, $T_{\rm co,I}$ and $T_{\rm co,II}$.}
\label{fig1}
\end{figure*}

The magnetotransport data obtained for LaRu$_{3}$Si$_{2}$ offer compelling evidence of the low-temperature charge-order transition. Magnetotransport \cite{Giraldo,Novak,XWei}, known for its sensitivity in detecting charge-order transitions, utilizes magnetoresistance (MR) as a measure of the mean free path integrated over the Fermi surface \cite{LDas}. This method is particularly adept at detecting changes in scattering anisotropy and Fermi surface reconstructions. In Figure 4$\bf{a}$, we present the MR under a perpendicular magnetic field in LaRu$_{3}$Si$_{2}$ across the temperature range of 10 K to 70 K. Within the primary $\frac{1}{4}$ charge-ordered state, MR remains zero, only emerging below the $\frac{1}{6}$ charge ordering temperature $T_{\rm CO,2}$. The MR exhibits an initial increase at $T_{\rm CO,2}$, with a faster increase below $T^{*}$, see Fig.~4$\bf{b}$. We note that not only does the absolute value of MR alter with decreasing temperature, but the overall profile of MR also undergoes modification as the temperature decreases. 


The Kohler plot, ${\Delta}{\rho}$/${\rho}_{H=0}$ vs (${\mu}_{0}H$/${\rho}{H=0}$)$^{2}$, displayed in Figure 4$\bf{c}$, is particularly noteworthy. The interesting observation in the Kohler plot is the singular behavior, where the MR data converge to a single line only below 15 K, in stark contrast to the pronounced temperature dependence evident at higher temperatures. Kohler's rule is a general principle stating that the normalized magnetoresistance, when plotted against a normalized magnetic field, should yield a universal curve for metals with simple Fermi surfaces. This principle tends to hold true even in materials with complex Fermi surfaces, provided there are no temperature-dependent changes in the details of the Fermi surface. The deviation from the Kohler's rule in LaRu$_{3}$Si$_{2}$ suggests a fundamental alteration in the electronic structure and scattering processes associated with the onset of secondary charge order, emphasizing its significant impact on the transport properties of the material. 


To further elucidate the electronic properties, Hall measurements were conducted, revealing a linear Hall resistivity at all temperatures. A linear Hall effect implies that transport is predominantly governed by a single type of charge carrier. In Figure 4$\bf{d}$, the Hall effect exhibits a pronounced temperature-dependence below $T_{\rm CO,2}$, with the sign of the Hall signal transitioning smoothly from positive (holes) to negative (electrons) across $T^{*}$. These findings contribute valuable insights into the charge-order-induced alterations in the electronic structure of LaRu$_{3}$Si$_{2}$. The sign reversal observed across the charge-ordering temperature in LaRu$_{3}$Si$_{2}$ parallels findings previously reported in 2H-NbSe$_{2}$ \cite{Wang2005} and cuprate high-temperature superconductors \cite{TailleferNature,TailleferPRB}. In those cases, this behavior was interpreted as a consequence of Fermi-surface reconstruction associated with the onset of a density-wave phase. It appears that in LaRu$_{3}$Si$_{2}$, the emergence of the secondary charge order similarly triggers a reconstruction of the Fermi surface, resulting in the appearance of an electron pocket. The similarities in the observed phenomena across different materials suggest a common underlying mechanism linked to charge order-induced alterations in the electronic structure.

\begin{figure*}[t!]
\centering
\includegraphics[width=1.0\linewidth]{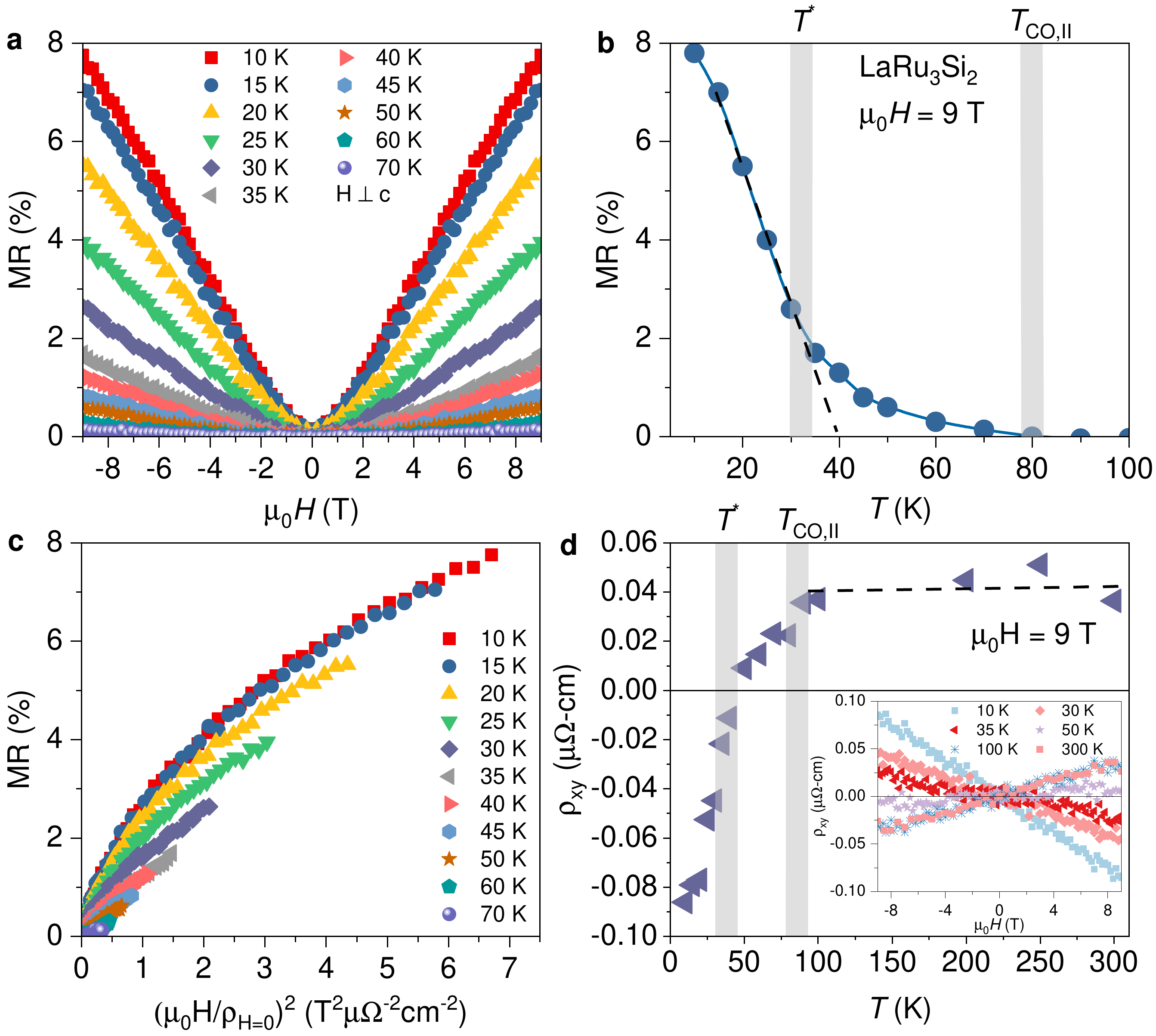}
\vspace{-0.5cm}
\caption{\textbf{Magnetotransport characteristics for LaRu$_{3}$Si$_{2}$.} 
$\bf{a,}$ The magnetoresistance measured at various temperatures across the charge ordering temperature $T_{\rm CO,2}$ ${\simeq}$ 80 K. 
$\bf{b,}$ The temperature dependence of the value of magnetoresistance at 9 T. $\bf{c,}$ Kohler plot, ${\Delta}$${\rho}$/${\rho}_{H=0}$ vs (${\mu}_{0}$$H$/${\rho}_{H=0}$)$^{2}$, of the magnetoresistance, plotted from field-sweeps at various temperatures. $\bf{d,}$ The temperature dependence of the value of the Hall resistance ${\rho}_{xy}$ at 9 T. Inset shows the Hall resistance, measured at various temperatures between 10 K and 300 K.}
\label{fig1}
\end{figure*}


\begin{figure*}[t!]
\centering
\includegraphics[width=1.0\linewidth]{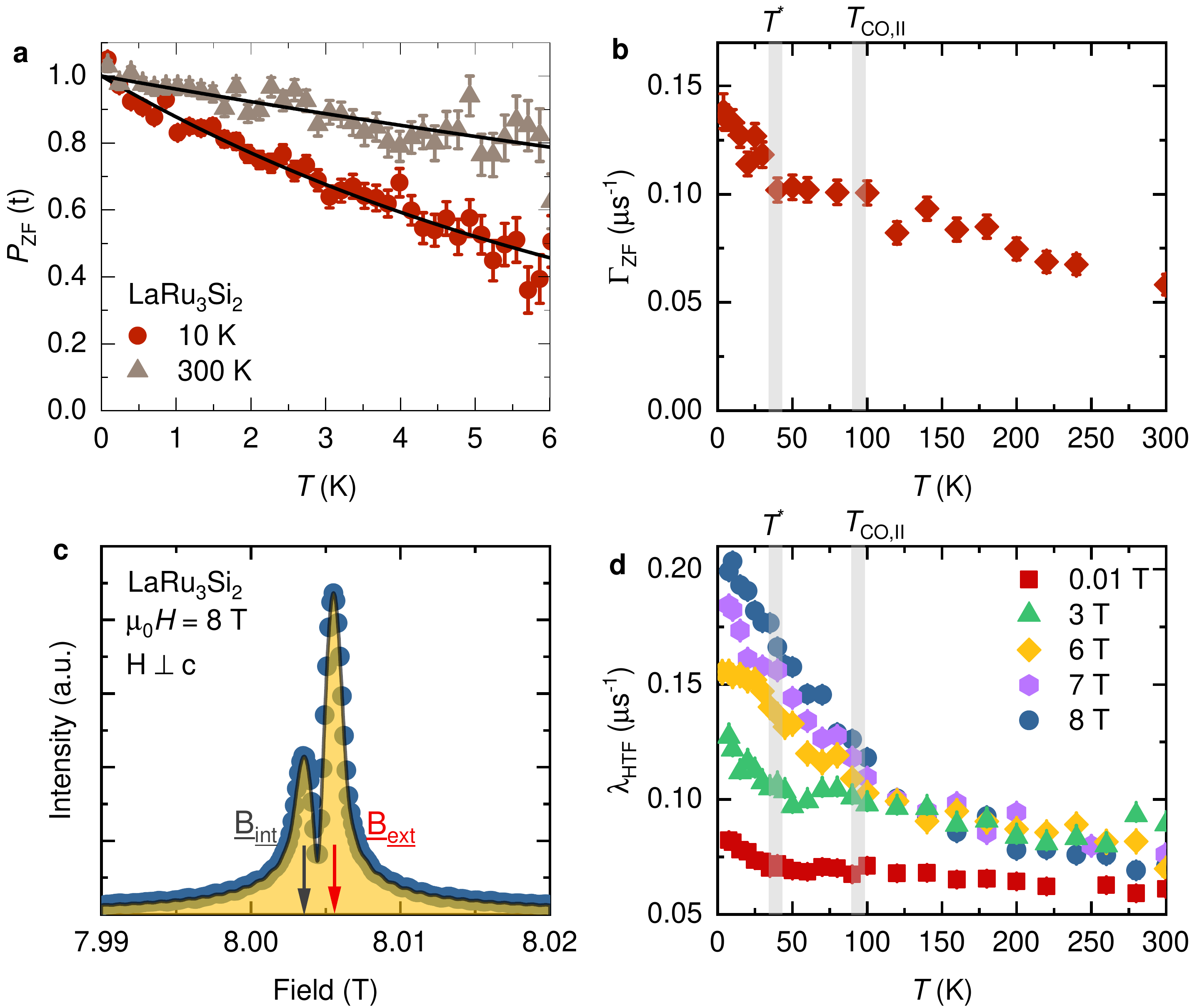}
\vspace{-0.5cm}
\caption{\textbf{Magnetic response of the charge order in zero-field and applying external magnetic fields in LaRu$_{3}$Si$_{2}$.} 
$\bf{a,}$ The ZF ${\mu}$SR time spectra, obtained at $T$ = 10 K and 300 K, all above the superconducting transition temperature $T_{\rm c}$. The solid black curves in panel $\bf{a}$ represent fits to the recorded time spectra, using the simple exponential function. Error bars are the standard error of the mean (s.e.m.) in about 10$^{6}$ events. $\bf{b,}$ The temperature dependences of the zero-field muon spin relaxation rate ${\Gamma}_{ZF}$, obtained over a wide temperature range. The error bars represent the standard deviation of the fit parameters. $\bf{c,}$ Fourier transform of the ${\mu}$SR asymmetry spectra for the single crystal of LaRu$_{3}$Si$_{2}$ at 5 K in the presence of an applied field of ${\mu}_{0} H = 8 $T. The solid line is a two-component signal fit. The peaks marked by the arrows denote the external and internal fields, determined as the mean values of the field distribution from the silver sample holder and from the sample, respectively. The short-time-window apodization function was used in the Fourier transform amplitude plot. $\bf{d,}$ Temperature dependence of the high transverse field muon spin relaxation rate $\lambda_{\rm HTF}$ for the single crystal of LaRu$_{3}$Si$_{2}$, measured under different $c$-axis magnetic fields. The vertical grey line marks the transition temperature below which the field effect was observed. The error bars represent the standard deviation of the fit parameters.}
\label{fig1}
\end{figure*}

Subsequently, motivated by the significant magnetoresistance at low temperatures in LaRu$_{3}$Si$_{2}$ and the magnetic response reported in the charge-ordered state of other kagome-lattice superconductors \cite{GuguchiaMielke,GuguchiaNPJ,GuguchiaRVS,KhasanovCVS,LiYu}, we conducted zero-field and high-field $\mu$SR (muon spin rotation) experiments on a single crystalline sample of LaRu$_{3}$Si$_{2}$. The zero-field ($ZF$)-$\mu$SR spectrum (Fig. 5$\bf{a}$) is characterized by a weak depolarization of the muon spin ensemble, indicating no evidence of long-range ordered magnetism in LaRu$_{3}$Si$_{2}$. However, it shows that the muon spin relaxation has a clearly observable temperature dependence. Since the full polarization can be recovered  by the application of a small external longitudinal magnetic field, $B_{{\rm LF}}$~=~5~mT, the relaxation is, therefore, due to spontaneous fields which are static on the microsecond timescale. The zero-field ${\mu}$SR spectra for LaRu$_{3}$Si$_{2}$ were fitted using the simple exponential function $P_{ZF}(t)$ = exp(-$\Gamma_{ZF} t$). Across the charge ordering temperature $T_{\rm CO,2}$ ${\simeq}$ 80 K, there is only a change in the slope of ${\Gamma}$. However, a more significant observation occurs as the temperature is lowered below $T^{*}$ ${\simeq}$ 35 K, where there is a notable increase in ${\Gamma}$. A possible reason for the change of the relaxation rate across $T^{*}$ could be an additional modulation of the lattice structure that slightly alters the nuclear positions around the muon \cite{Bonfa}. However, a rough order of magnitude estimate yields that the structural distortions of the order of 0.1{\AA} for the atoms closest to the muon would be needed to explain the observed effect in the second moment of the measured field distribution. This is a large effect that has not been seen by X-ray diffraction experiments. Therefore, we can dismiss the structural distortion being the origin for the increase of the relaxation rate. Most importantly, our high-field ${\mu}$SR results presented here definitively prove that there is indeed a strong contribution of electronic origin to the muon spin relaxation below the charge-ordering temperature (see below). Therefore, we interpret our ZF-$\mu$SR results as an indication that there is an enhanced width of internal fields sensed by the muon ensemble below $T^{*}$ ${\simeq}$ 35 K. The increase in ${\Gamma}$ below $T^{*}$ is estimated to be ${\simeq}$ 0.035~${\mu}$$s^{-1}$, which can be interpreted as a characteristic field strength ${\Gamma}$/${\gamma_{\mu}}$ ${\simeq}$ 0.4~G.

To corroborate the zero-field ${\mu}$SR results presented above, a comprehensive set of high-field ${\mu}$SR experiments was conducted \cite{Sedlak}.
Figure 5$\bf{c}$ illustrates the probability field distribution measured at 5 K in a magnetic field of 8 T applied perpendicular to the crystallographic $c$ axis. Throughout the investigated temperature range, the ${\mu}$SR signals were characterized by two components: a signal with fast relaxation ($\lambda_{\rm HTF}$ ${\simeq}$ 0.428(3) ${\mu}s^{-1}$) and another with slower relaxation (0.05 ${\mu}s^{-1}$). The fast relaxation signal, predominant in the ${\mu}$SR spectrum, arises from muons stopping within the sample and indicates a bulk response. Fig. 5$\bf{d}$ depicts the behavior of the relaxation rate in a magnetic field of 0.01 T applied parallel to the $c$ axis, resembling the temperature dependence observed in zero-field conditions. At higher fields (3 T, 6 T, 7 T, and 8 T), the temperature dependence of the relaxation rate varies, with a clear and monotonous increase below the onset temperature $T_{\rm CO,2}$. This suggests that the low-temperature relaxation rate in magnetic fields is predominantly influenced by electronic/magnetic contributions, as nuclear contributions are not susceptible to external field enhancement. The absolute increase of the relaxation rate between the onset of charge order $T_{\rm CO,2}$ and the base temperature in 8 T is ${\Delta}\lambda_{\rm HTF}$ ${\simeq}$ 0.1 ${\mu}s^{-1}$, which is smaller than in other kagome-lattice superconductors such as KV$_{3}$Sb$_{5}$ \cite{GuguchiaMielke} and ScV$_{6}$Sn$_{6}$ \cite{GuguchiaSc166}. Since the onset of the weak magnetic response aligns with the temperature $T_{\rm CO,2}$ of ($\frac{1}{6}$,~0,~0) charge order, this indicates a strong intertwining between magnetic and charge channels. Additionally, a weak but non-negligible field effect on the relaxation rate is observed above $T_{\rm CO,2}$, extending up to at least 300 K. This effect may be associated with the high-temperature charge-ordered state, and further experiments above 300 K are crucial for a more comprehensive understanding.

In this paper, we have presented three main findings: (1) Charge order with a propagation vector of ($\frac{1}{6}$, 0, 0) is identified in the single crystal of LaRu$_{3}$Si$_{2}$ below a critical temperature $T_{\rm CO,2}$ $\simeq$ 80 K, coexisting with a previously reported room-temperature primary charge order ($\frac{1}{4}$, 0, 0). The charge order persists into the superconducting state, highlighting the intricate interplay between superconductivity and distinct charge-order phenomena.  (2) Magnetotransport measurements reveal zero magnetoresistance within the primary charge-ordered state and the appearance of magnetoresistance below $T_{\rm CO,2}$. The Kohler plot analysis, showing a violation of Kohler's rule, suggests a Fermi surface reconstruction induced by charge order. Anomalies in the Hall resistivity and its sign reversal across the charge-ordering temperature further indicate a complex interplay of electronic states. The reconstruction of the Fermi surface is a fundamental aspect of such phenomena, wherein the charge order induces changes in the distribution of electronic states near the Fermi level. This restructuring can lead to anomalous transport properties, such as the observed temperature-dependent Hall resistance and its sign reversal. By drawing parallels to observations in other materials such as cuprate high temperature superconductors, the study of LaRu$_{3}$Si$_{2}$ provides valuable insights into the universal aspects of charge order effects on electronic behavior, contributing to the broader understanding of quantum materials. (3) The $\mu$SR experiments conducted in both zero-field and high-field conditions have yielded noteworthy insights into the properties of LaRu$_{3}$Si$_{2}$. Below $T^{*}$, approximately 35 K, there is a notable enhancement in the internal field width sensed by the muon ensemble. Additionally, a pronounced field-induced enhancement of the relaxation rate is observed below $T_{\rm CO,2}$, which is around 80 K. These observations collectively indicate time-reversal symmetry breaking within the secondary charge-ordered state.

The debate about the origins \cite{Mazin2008} of the various forms of charge order found in Kagome systems focuses on whether they are caused by electronic interactions or phononic effects. Electronically, there is significant interest in how sublattice interference and Q vectors connect the van Hove singularities (vHS) \cite{Kiesel,MDenner}. The proximity of VHS to the Fermi level is key to the unusual superconductivity and 2${\times}$2 charge order in the $A$V$_{3}$Sb$_{5}$ ($A$=K, Rb, Cs) compounds  \cite{YJiang}. This 2${\times}$2  charge order pattern is also observed in the magnetic Kagome lattice of FeGe \cite{JiaxinPRL,XTeng}, exhibiting similarities with the charge order found in Kagome superconductors $A$V$_{3}$Sb$_{5}$. In ScV$_{6}$Sn$_{6}$ \cite{Arachchige}, the band structure also shows vHS near the Fermi level, but its charge order differs from the $A$V$_{3}$Sb$_{5}$ series. In ScV$_{6}$Sn$_{6}$ \cite{Arachchige}, the primary driver for charge order is believed to be phonons, with the leading distortion involving an out-of-plane modulation of Sn and Sc sites~\cite{SCao,HHu,YongMing,Blanco2023}. This is in contrast to the in-plane Kagome breathing mode in $A$V$_{3}$Sb$_{5}$.  The Kagome superconductor LaRu$_{3}$Si$_{2}$ also has vHS in its band structure, relating to Ru-$dz^{2}$ orbitals near the Fermi level  \cite{GuguchiaPRM}. Its charge order \cite{GuguchiaPlokhikh} involves the in-plane displacement of Si atoms and out-of-plane displacements of some Ru atoms, differentiating it from both $A$V$_{3}$Sb$_{5}$ and ScV$_{6}$Sn$_{6}$. Additionally, the superconductor LaRu$_{3}$Si$_{2}$ exhibits a distinctive three-dimensional character, both in its band structure and in its isotropic superconducting properties. This contrasts with the more two-dimensional nature of the $A$V$_{3}$Sb$_{5}$. Despite these variations in charge-order structures and their formation mechanisms (electronic and phononic) in different Kagome superconductors, they commonly exhibit indications of time-reversal symmetry breaking in the charge-ordered state \cite{GuguchiaMielke,GuguchiaNPJ,GuguchiaRVS,KhasanovCVS,GuguchiaSc166}. This suggests a prevalent occurrence of time-reversal symmetry-breaking in charge-ordered Kagome lattices, transcending specific fermiologies. LaRu$_{3}$Si$_{2}$  is unique among Kagome superconductors for its coexistence of multiple charge orders, each with distinct magnetic responses. It might be conjectured that  the high-temperature transitions at $T_{\rm str}$ and $T_{\rm CO,1}$  are phonon-driven, while for the low-temperature transition at $T_{\rm CO,2}$, electronic interactions are relevant. These findings deepen our understanding of how lattice and electronic effects interact to cause charge order instabilities in these complex systems. The presence of various orders at different temperatures might relate to orbitally selective ordering, adding complexity to the understanding of these materials' electronic and magnetic properties and highlighting distinctiveness of LaRu$_{3}$Si$_{2}$ in the Kagome superconductor family.

\section{METHODS}

%

\textbf{X-ray diffraction}: X-ray diffraction was performed at ID27, ESRF using monochromatic x-rays with a wavelength of 0.3738 ${\AA}$ and a spotsize of 0.6x0.6 ${\mu}$m$^{2}$. The sample was mounted on a membrane driven diamond anvil cell with 70 degrees angular aperture designed at ESRF. Helium was used as pressure transmitting medium and the applied pressure was measured by Ruby fluorescence. The diamond anvil cell was mounted into a Helium flow cryostat (ESRF). The applied pressure was kept constant to 0.3 GPa during cooling by adapting the membrane pressure. X-ray diffraction was collected with a Eiger2 CdTe detector (DECTRIS AG, Baden-Daettwil, Switzerland) in shutterless mode and continuous rotation over 64 degrees with readout every 0.1 degree. CrysAlisPro (Rigaku) was used for data reduction and reciprocal space reconstructions. Our prior measurements using muon spin rotation revealed that applying pressures up to 2.2 GPa does not change the physical properties of this material. This suggests that a pressure of 0.3 GPa can be regarded as the ambient pressure for LaRu$_{3}$Si$_{2}$.\\

\textbf{Magnetotransport}: The magnetoresistance of the single-crystal LaRu$_{3}$Si$_{2}$ was assessed using the physical property measurement system (PPMS-9, Quantum Design) employing the conventional four-probe technique. Four Pt-wires (0.0254 mm diameter) were affixed to the single crystal using silver epoxy, forming a bar-shaped specimen after polishing. A constant electrical current of 1 mA was applied, and the magnetic field was directed along the crystallographic $a$- and $c$-axes, a configuration validated through Laue measurement.\\ 

\textbf{${\mu}$SR experiment}: In a ${\mu}$SR (muon spin rotation) experiment, nearly 100${\%}$ spin-polarized muons ${\mu}^{+}$ are introduced into the sample one at a time. These positively charged ${\mu}^{+}$ particles thermally stabilize at interstitial lattice sites, effectively serving as magnetic microprobes within the material. In the presence of a magnetic field, the muon spin undergoes precession at the local field $B_{\rm \mu}$ at the muon site, with a Larmor frequency ${\nu}_{\rm \mu}$ given by $\gamma_{\rm \mu}$/(2${\pi})$$B_{\rm \mu}$, where $\gamma_{\rm \mu}$/(2${\pi}$) represents the muon gyromagnetic ratio and is equal to 135.5 MHz T$^{-1}$.\\

Zero field (ZF) and transverse field (TF) $\mu$SR experiments on the single crystalline sample of LaRu$_{3}$Si$_{2}$ were performed on the GPS instrument and high-field HAL-9500 instrument, equipped with BlueFors vacuum-loaded cryogen-free dilution refrigerator (DR), at the Swiss Muon Source (S$\mu$S) at the Paul Scherrer Institut, in Villigen, Switzerland. Large single crystal piece was used for these measurements. 
The crystal, with dimensions 7$\times$1.7$\times$0.7~mm$^{3}$, was affixed to a 5 mm circular silver sample holder using a small droplet of GE varnish during high-field experiments. The magnetic field was applied perpendicular to the crystallographic $c$-axis. The crystal was mounted such that the $c$-axis of it is perpendicular to the muon beam. Using the ''spin rotator'' at the ${\pi}$M3 beamline, muon spin was rotated (from its natural orientation, which is antiparallel to the momentum of the muon) by $44.5(3)^{\circ}$ degrees with respect to the $c$-axis of the crystal. So, the sample orientation is fixed but the muon spin was rotated. The rotation angle can be precisely determined to be $44.5(3)^{\circ}$ by measurements in weak magnetic field, applied transverse to the muon spin polarization. Zero field and high transverse field $\mu$SR data analysis on single crystals of LaRu$_{3}$Si$_{2}$ were performed using both the so-called asymmetry and single-histogram modes \cite{Suter,Yaouanc}. The experimental data were analyzed using the MUSRFIT package \cite{Suter}.\\

\section{Acknowledgments}~
The ${\mu}$SR experiments were carried out at the Swiss Muon Source (S${\mu}$S) Paul Scherrer Insitute, Villigen, Switzerland. X-ray diffraction experiments were performed at ID27 at European Synchrotron Radiation Facility (ESRF), France. Z.G. acknowledges support from the Swiss National Science Foundation (SNSF) through SNSF Starting Grant (No. TMSGI2${\_}$211750). Z.G. acknowledges the useful discussions with Robert Scheuermann. I.P. acknowledges support from Paul Scherrer Institute research grant No. 2021${\_}$0134. \\

\textbf{Author Contributions}
Z.G. conceived and supervised the project. Growth of the single crystal LaRu$_{3}$Si$_{2}$: H.N., and S.N.. 
Single crystal X-ray diffraction experiments, analysis and corresponding discussions: B.W., C.M.III, I.P., V.S., S.S., J.N.G., G.G., J.K., I.B., J.C., E.P., D.J.G., and Z.G.. Magnetotransport experiments: V.S., M.B., and Z.G.. Magnetization experiments: C.M.III, M.M., D.J.G., and Z.G. 
Muon spin rotation experiments and the corresponding discussions: C.M.III, V.S., J.N.G., D.D., S.S.I.,  M.H.F., T.N., J.-X.Y., M.Z.H, H.L., R.K., and Z.G..
Figure development: V.S., C.M.III, and Z.G.. Writing of the paper: Z.G. with contributions from all authors. All authors discussed the results, interpretation and conclusion.\\


%
%

\end{document}